# Quantum Calculations Show Caution Is Needed In Interpreting Methanethiosulfonate Accessibility Experiments On Ion Channels

Alisher M. Kariev and Michael E. Green

Department of Chemistry

City College of New York

New York NY 10031

green@sci.ccny.cuny.edu

Standard models of ion channel voltage gating require substantial movement of one transmembrane segment, S4, of the voltage sensing domain. Evidence comes from the accessibility to external methanethiosulfonate (MTS) reagents of the positively charged arginine residues (R) on this segment. These are first mutated to cysteines (C), which in turn react with MTS reagents; it is assumed that the C is passively carried in the S4 movement, becoming accessible on one side or the other of the membrane. However, the Rs were salt bridged to negatively charged residues on other transmembrane segments, or hydrogen bonded, while C reacts as a negative ion. The space available for MTS is fairly close to the difference in volume between the large R residue and much smaller C, so the MTS is not severely sterically hindered. A reagent molecule can reach a C side chain; the C can react if not repelled by a negative charge from the amino acid to which the R had been salt bridged. Nearby protons may also make reaction possible unless the C itself is protonated. Therefore interpretation of the C substitution results requires reconsideration. To test the idea we have done quantum calculations on part of a mutated S4 and the nearby section of the channel. The mutation is R300C of the 2A79/3Lut structure, a mutation that would be done to test MTS reagent access; there is a large cavity where the R is replaced by C. Two quantum calculations show a substantial difference in the structure of this cavity with 2 water molecules compared to 4. This suggests that the structure, and presumably reaction probability, could depend on water molecules, very likely also protons, in or near the cavity that the R300C mutation produces.

INTRODUCTION: Ion channels, which are among the most important proteins in cell membranes, come in various forms, of which voltage gated channels are a major subset, responsible for much of the activity of nerves and other excitable tissue[1]. These channels respond to membrane depolarization by producing a small capacitative current that precedes the opening of the channel, called the gating current. Most models of gating attribute this to the motion of arginine residues on one transmembrane (TM) segment, S4, in the voltage sensing domain (VSD). The models exist in various forms: *1)* the segment corkscrews upward (by convention the extracellular direction is "up") upon membrane depolarization (helical screw model[2, 3]); *2)* S4 moves accompanied by part of a neighboring segment (paddle model)[4] *3)* the channel pushes the ion through as a transporter would [5] *4)*variations on similar themes. The evidence is, in detail, not adequate to confirm any of these models. However, in every version, the gating current is produced by a finite physical displacement of the positive charges on arginines (rarely lysines).

We have proposed a model in which instead of this physical motion of protein, the gating current consists of the motion of protons part way through the membrane; it is known that substituting a histidine for the terminal arginine allows a proton current to go all the way through the membrane, but substituting the histidine for a neighboring hydrophobic residue does not allow such a current, so that it is a current of moving protons, not protons being carried by the physical motion of the histidines[6, 7]. Thus we can be sure that the protons can move along the path we propose; presumably with the original arginine in place, the proton is stopped and no continuous current through the VSD is possible in the native structure. The $H_v1$ channel, a proton channel strongly resembling the VSD, leads to a similar conclusion[8, 9]. In order to take this alternate model seriously it is necessary to reconsider the strongest evidence for the physical motion of the S4 segment of the VSD.

Much evidence for the standard models is based on the Substituted Cysteine Accessibility Method (SCAM)[10]. Here a residue is mutated to cysteine, which in turn reacts, if it can be reached, with a methanethiosulfonate (MTS) reagent. For the voltage gated ion channels, the residues that are mutated are, in most cases, the arginines of S4 of the VSD of a voltage gated ion channel, and what we are saying applies specifically to this case. We are not discussing the many other applications of the SCAM method.

*Major evidence in the argument for the physical movement of the S4 is based on apparent accessibility of the cysteine to the MTS reagents. This changes with the state of the channel, being different in the open state and the closed state. In the open state, there is more apparent access near the "top" (extracellular end), in the closed state, near the bottom.* In these models, access in the open state is at the top because the S4 has moved up with respect to the other TM segments and the membrane, exposing the topmost two (usually two) arginines as indicated by the reaction of the cysteines that have replaced these arginines. When the channel is closed there is only one available arginine at the top, but at least one more, often two more, at the bottom. However, it must be understood that not arginines but cysteines placed there by R→C mutations are reacting, and these need not behave as do arginines. It is a seemingly natural assumption that the differential accessibility of cysteines is a consequence of the physical motion of the S4. If, as we are suggesting here, the S4 is stationary, while protons move, only the interpretation of the access results must be different, with no change in the experimental findings. Here, we suggest how this is possible.

THE PROBLEM IN THE STANDARD INTERPRETATION: The main points in the qualitative argument for reinterpretation of the results as other than physical movement of the S4 segment are as follows:

1) The cysteine, in order to react, must lose an $H^+$, leaving the cysteine anion, which is the reactive species[11]. Therefore the reactive form constitutes a charge reversal mutation, not at all a conservative mutation.

2) At least two arginines are salt bridged to aspartates or glutamates in the S2 and S3 TM segments of the VSD in the most studied $K^+$ channel, Shaker, and closely related channels. Thus there are negative charges in the immediate vicinity of the new negative charge on the reactive form of the cysteine. Coulomb forces could then cause the cys sulfur to fold back to the backbone, and be unavailable to react.

3) The cysteine is much smaller than the arginine (see Fig. 1), leaving room for the MTS reagent, so that sterically, access to the vicinity of the cys is not so difficult, although reaction remains difficult, as the reactive moiety, essentially the $S^-$, is likely to be hidden away from the MTS reagent, repelled by the neighboring aspartate or glutamate residue. Fig. 1b and Fig. 1c show a large gap, or cavity, where the arginine had been in the WT (Fig. 1a), when a cys is present instead of R300.

4) Interactions with other neighbors also change drastically. In the 2A79/3Lut structure of the channel, there is a cysteine, C229, from S2, and a serine, S176, from S1, that are both close neighbors of R303. R297 has a threonine in the vicinity, and this shifts in R300C. Whether these form a sort of "button" that ties down the arginine or not, it should complicate understanding of the effect of an R→C mutation at location 300, since it would not even be clear which cysteine reacted; it would be surprising if such close neighbors could be neglected. In addition, R300 has a strong salt bridge to a glutamate, E226; an amino N of the R300 side chain is only 2.71 Å from a glutamate O atom, with an H near the line between. There is nothing comparable in the R300C mutant. This is the strongest (inferred from bond length) salt bridge of all the three arginines in the calculation. In WT, R297 is close to 5 Å from its counterion, E183, indicating a salt bridge that is not as strong, and probably held apart by the presence of R300. In the mutant with four waters (Fig. 1c), the distance drops to ≈4.5 Å with water present. Water can make a salt bridge stronger, especially as the salt bridge need not be ionized without water [12]. We have done quantum calculations, optimizing the structure, with two water molecules and with four water molecules in or near the cavity formed by the mutation. These can be compared to the WT structure. In Fig. 1, the calculated mutants (two and four water molecule cases) are compared to the X-ray structure of the wild type. Also in the calculation with two waters (Fig. 1b), we see that one of the waters appears in the R297-E183 salt bridge. This is still a fairly strong interaction, albeit less so than the one that included R300. A putative S4 shift could not replace one salt bridge with another without a significant energy penalty.

5) A secondary consideration: even if we were to argue that the comparison is R303 to R297, with the R300 position always occupied in a large S4 shift, the energies should differ considerably, and multiple hydrogen bonds would be affected, requiring an even larger energy of activation than the energy difference between states with shifted salt bridges. Going through this in detail would carry us too far afield from the calculation we were able to do, optimizing geometry, but some consequences of motion can be seen from just inspecting the structures. There is also a salt bridge of R303 to the same E226, but at a slightly larger distance, two atoms being approximately 4 and 5 Å away. Again, this is not the same type of salt bridge as the R300-E226 case, and again a shift of salt bridges would produce an energy penalty in one direction. Such a shift would also produce a local structural rearrangement that would be more severe than an exchange of salt bridges, as other amino acids, such as Y266, which participates in a hydrogen bond network that would be disrupted, would have to rearrange as well to allow any S4 displacement. Other amino acids in the neighborhood likely to be involved in the hydrogen bonding include C229 and S176. The water at the edge of this region would also have to move and rearrange its hydrogen bonds. Thus S4 displacement would mean a very large alteration in local bonding, involving several salt bridges and hydrogen bonds. In the WT structure, water would not be present as R300 fills the space. In the mutant, any motion would require significant water rearrangement as well, another major difference between WT and mutant.

6) The MacKinnon group experiment, in which an MTS – cys reaction was used to label the channel with biotin with variable chain length [4] requires special consideration, in that the mutations were not limited to arginines. However, in almost every case, the cys was substituted for amino acids with a larger side chain; in principle, these could be different than the arginine mutations, and a special calculation could be done to test these. However, given that these experiments depend on cys substitution for a larger amino acid, it is to be expected that considerations similar to those applicable to the arginine mutations would apply here. Some mutations are in fact R→C, while others are mutations from neutral amino acids. As charge reversal is not involved, it may not be as drastic a change as the arginine case. Nevertheless the changes in local geometry, and for that matter charge, are not easily predictable, and the access question cannot be easily addressed; it need not be as simple as a geometrical length of the fully extended spacer chain from the membrane surface. Folds may occur in the cavity, or other distortions, including partial charge transfers, are possible, since the internal geometry of the VSD has been altered by the mutation. Reaction probability may also be affected by the location of protons, so that reaction/failure to react may still be state dependent without geometric alterations. We have not done the calculation here, nor have we gone through the mutations one at a time to check for patterns, so the question remains slightly open. However, we do not see how a simple distance from the surface of the membrane can be inferred from the results of these experiments, any more than from the S4 R→C mutations. Even though the linkers appear to establish a length scale, the question of whether reaction has occurred prevents a direct interpretation of the results. The technique is ingenious but not conclusive.

There is an alternate way of dealing with the negative potential produced by the carboxylic acid to which the arginine had been salt bridged, and the negatively charged reactive form of the cysteine, if the gating current consists of protons. In that case, the negative potential near the two amino acids and the MTS reagent can be sufficiently neutralized that the cys residue can unfold and react with the MTS reagent. We have calculated the structure of the cys/asp/neighbor structure with no added protons, but with two and four water molecules, as shown in Fig. 1b and 1c; there are too many variables to deal with in cases in which the structure is not known, as would be the case with added protons, for example. The reaction itself has not been calculated for similar reasons. Qualitatively we can observe that unless a proton simply neutralizes the cys, the presence of the favorable electric field in the presence of the protons would answer the question of the state dependence of the apparent accessibility. Because the proton would insert into a multiresidue network (several could be imagined, depending on local rearrangements), it is likely that the resulting "acid" is not a single residue, like the cysteine that has been substituted, and pK would be very difficult to predict or calculate. The difference in the structure with two, as against four, water molecules, already shows how large such an effect might be, even without charge. Any interpretation of the reaction with MTS reagents must include accounting for the major rearrangement of hydrogen bond networks such a reaction would require.

The first five points show that interpreting the results of a R→C mutation requires further discussion; these mutations should not at this time be regarded as offering unambiguous evidence of any mechanism, whether physical movement of some particular form, or proton rearrangement. A charge reversal mutation, with major change in volume of the residue, gives serious reason to question an interpretation that requires a straightforward reaction with no significant strictly local conformational change affecting reaction probability. Given the possibilities of local rearrangements, assuming the cys residue is carried passively with an assumed S4 overall motion remains questionable. The discussion in point 6, concerning the Mackinnon method, leads to a similar conclusion.

We have done an ab initio (HF/6-31G**) optimization of the region of one of the standard mutations, including the wild type, the mutant with two water molecules, and the mutant with four water molecules (Fig. 1). We see that the structure is in fact greatly altered, with a large cavity appearing in the place of the arginine, when only a cysteine is present to replace it. It is clear that this mutation causes major changes in the structure, and in particular in the availability of space for a reagent, like MTS, to reach the cysteine. Given the difference between two and four water molecules, the state dependence that is observed in substitution reactions should not be easily interpreted as implying that the cysteine has emerged at the membrane surface.

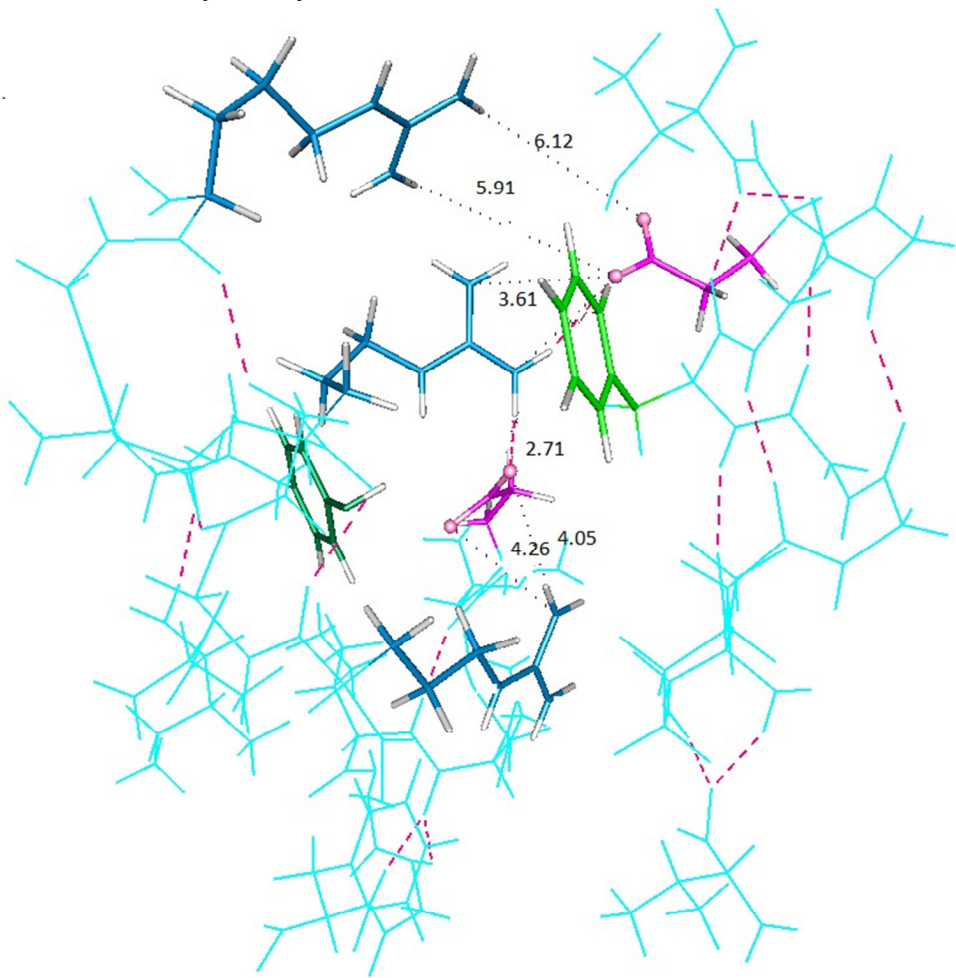

Fig. 1A

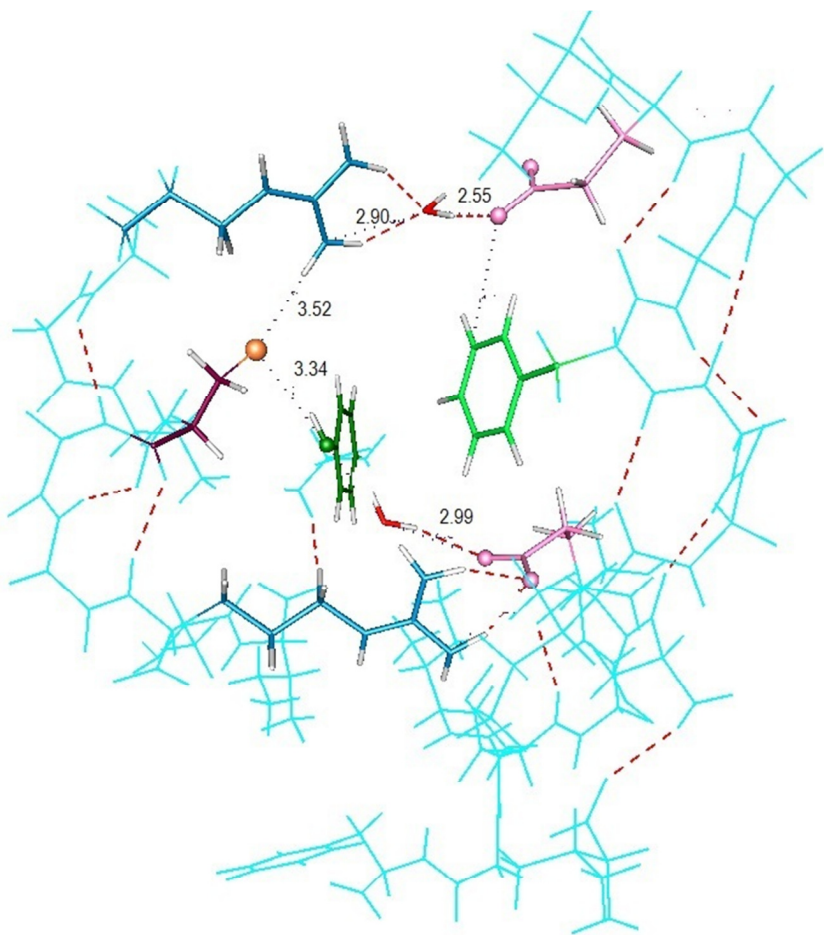

Fig. 1B

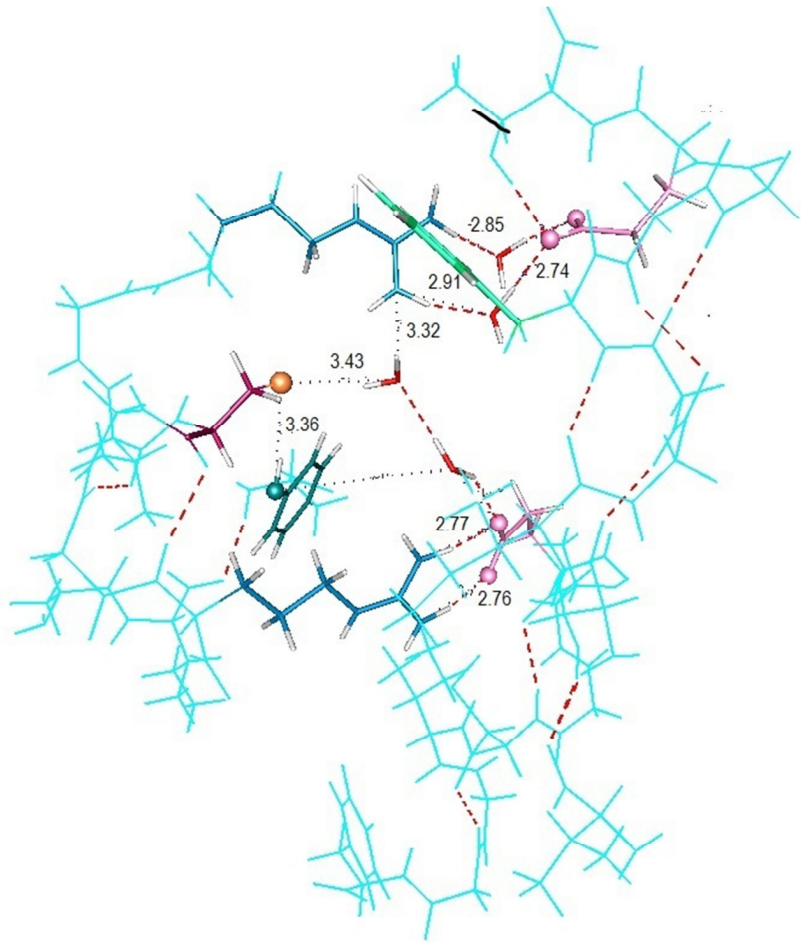

FIGURE 1: The central part of the voltage sensing domain, from R297 to R303 on the S4 TM segment, and nearby residues on S1, S2, S3. Most of the protein is shown as thin blue lines. Arginines are heavy blue lines, glutamates heavy purple lines with carboxyl oxygens shown as purple spheres, tyrosine (Y266) is dark blue hydroxyl oxygen indicated as a sphere, phenylalanine F233 green, and in B and C, C300 is red, with the sulfur a large orange sphere on the left side of the figure. The water molecules are red (oxygen) and white (hydrogen) rods. Hydrogen bonds are dashed lines, and certain distances are indicated in Angstroms.

A) Wild type: Observe the central arginine stretching across the space between S4 and the rest of the VSD. There is essentially no space for an external reagent to enter the space. It would be difficult to place even a water molecule in the center of the region. Certain key distances are shown on the diagram. It is apparent that the only place for water molecules is the top salt bridge which is clearly fairly weak, with approximate distances 5 to 6 Å.

B) In this case, the R300C mutation is shown, with two water molecules. The water molecules both form hydrogen bonds, one with the R297 and corresponding glutamate (E183), with the oxygen of the water equally hydrogen bonded to two hydrogens of the arginine; the other water is near R303. The sulfur of the cysteine of the R300C mutation is doubly hydrogen bonded, to Y266 and to R297, and held to the side of the cavity; the S to nearest R297 atom distance is 3.56 Å, suggesting a strong bond. F233 rotates into the cavity.

C) This shows the R300C mutation with four water molecules. There are two water molecules now bridging R297 to E183, and two in the cavity center, one hydrogen bonded to the cysteine sulfur and to the other water molecule, which stretches across to the glutamate (E226) on the other side of the cavity. The sulfur atom now points into the cavity, with its second hydrogen bond to tyrosine (Y266). It appears much more available to a reagent in the cavity. The hydrogen bond network near Y266 and C300 has also shifted considerably compared to the two water case. The F233 ring has folded sharply out of the way (with dihedral angle defined by ring C4, ring C1 (the link to the atoms toward the backbone) and the next two atoms of the side chain toward the backbone of +171.3°, compared to -99.1° in the two water case).

CALCULATION: The calculation allows testing a key assertion made above. We can optimize the structure of an R→C mutant to see what happens to the sulfur atom, and test whether it does fold away from the negatively charged carboxyl group. This is shown in Fig. 1b and 1c, using the R300C mutant, and the qualitative assertion that we made is largely validated by the computation, albeit not exactly as we had originally expected, since Y266 plays a significant role that involves hydrogen bonding instead of electrostatic repulsion. R297, R300, R303, and their neighbors in the 2A79 K$_v$1.2 structure, plus the glutamate E226 from the neighboring S2 transmembrane segment, and three other amino acids, including a neighboring cysteine, C229, are included, along with Y266 and S176. Essentially, every amino acid that has a side chain pointing into the cavity, or toward the arginine, in the X-ray structure, is included in full. The connecting amino acids, with side chains pointing away in the X-ray structure, have only their backbone atoms included. In total there are 380 protein atoms, plus the 6 or 12 water atoms, in the calculation. The arginines above and below have glutamate salt bridges, albeit less strong than that of R300, as discussed above. Fig. 1a shows the WT with the X-ray structure, Fig. 1b the R300C mutant with 2 molecules of water in the computation, Fig. 1c the same with 4 waters, both optimized using Gaussian09[13] at HF/6-31G** level. In the mutant, we start with only the substituted cys negatively charged, not the native C229. Distances are shown in the figure, and it can be seen that the neighbors are quite close; it is marginal whether there is room for a water molecule in the WT in the central space occupied by R300, although there would be water in the R297-E183 salt bridge. While the original (WT) form is nearly space filling, the mutant R300C leaves a large cavity, into which we have placed the water molecules. This space presumably could be filled by an MTS reagent, if the potentials allowed. However we have an important observation in the optimized structure that the cysteine sulfur moves toward the Y266, forming a sort of hydrogen bond with the tyrosine hydroxyl hydrogen, with distances of 3.34 or 3.36 Å. This is a way to keep the sulfur from reacting we had not considered, but would be effective. It is clear that there is a strong probability that the substitution of cys for arg is not a benign mutation that leaves structure unchanged. Other cys substitutions could produce less drastic effects, and the calculation presented here would give a different result, in detail, for another mutation. Any of the mutations, however, would be seriously disruptive of local structure. Also, for any other mutation, the strong R300-E226 salt bridge would need to be disrupted if S4 were to slide down as the channel closed (the X-ray structure shows the open form). For a test of this interpretation it would be very useful to have an X-ray structure of the cysteine mutant without any MTS reagent to see how the cysteine is arranged; so far, this has not been done.

CONCLUSIONS: 1) The mutation of an arginine to cysteine is not a harmless substitution which can be assumed to allow the passive transport of the substituted cysteine if there were an S4 displacement. Standard models require cys to follow S4 to essentially the same position the arginine would have had when it moves, in whatever manner that model requires. That is, all such models assume that there is conformational change that is unaffected by the substitution. We see here that the results of the substitution, followed by MTS reaction, cannot be used directly to infer the nature of a conformational change, or even whether any such change occurs at all.

2) The difference in volume of the cys and arg is sufficient for at least two water molecules, or for the head group of an MTS reagent. Including the boundary of the cavity allows up to at least four water molecules. The structure shifts when two more water molecules are added. Water can clearly affect the reactivity of the substituted cys; nearby protons can be expected to do at least as much.

3) The cys, when substituted for R300 and in its reactive state (negatively charged) forms a hydrogen bond with neighboring Y266. It moves slightly, but sufficiently, in the optimization, thus becoming apparently out of reach of the MTS reagents. This is one of several local, small, conformational shifts that the mutation induces. The state dependence cannot easily be predicted.

4) Therefore it is not possible to conclude that MTS reaction with a substituted cys is strong structural evidence of S4 motion.

ACKNOWLEDGMENT: Research carried out in part at the Center for Functional Nanomaterials, Brookhaven National Laboratory, which is supported by the U.S. Department of Energy, Office of Basic Energy Sciences, under Contract No. DE-AC02-98CH10886.